\documentclass[aps,prd,preprint,draft,showpacs]{revtex4}
\usepackage{graphicx,epsf,amssymb}
\newcommand{\be}{\begin{equation}} 
\newcommand{\ee}{\end{equation}}

\newcommand{\bea}{\begin{eqnarray}} 
\newcommand{\eea}{\end{eqnarray}} 
\newcommand{\Tr}{{\rm Tr}}

\newcommand{\NeqFour}{{\cal N} =4}

\newif\ifdraft
\drafttrue
\newif\ifpreprint
\preprinttrue

\def\Sect#1{Section~{\ref{#1}}}
\def\sect#1{section~{\ref{#1}}}

\def\fig#1{fig.~{\ref{#1}}}

\def\tree{{\rm tree}}
\def\pol{\varepsilon}
\def\Tr{\, {\rm Tr}}

\def\NeqFour{{\cal N}=4}
\def\NeqOne{{\cal N}=1}

\def\spa#1.#2{\left\langle#1\,#2\right\rangle}
\def\spb#1.#2{\left[#1\,#2\right]}
\def\sand#1.#2.#3{%
\left\langle\smash{#1}{\vphantom1}^{-}\right|{#2}%
\left|\smash{#3}{\vphantom1}^{-}\right\rangle}
\def\sandp#1.#2.#3{%
\left\langle\smash{#1}{\vphantom1}^{-}\right|{#2}%
\left|\smash{#3}{\vphantom1}^{+}\right\rangle}
\def\sandpp#1.#2.#3{%
\left\langle\smash{#1}{\vphantom1}^{+}\right|{#2}%
\left|\smash{#3}{\vphantom1}^{+}\right\rangle}
\def\sandpm#1.#2.#3{%
\left\langle\smash{#1}{\vphantom1}^{+}\right|{#2}%
\left|\smash{#3}{\vphantom1}^{-}\right\rangle}
\def\sandmp#1.#2.#3{%
\left\langle\smash{#1}{\vphantom1}^{-}\right|{#2}%
\left|\smash{#3}{\vphantom1}^{+}\right\rangle}
\def\sandmm#1.#2.#3{%
\left\langle\smash{#1}{\vphantom1}^{-}\right|{#2}%
\left|\smash{#3}{\vphantom1}^{-}\right\rangle}
\def\spab#1.#2.#3{\sandmm#1.#2.#3}
\def\spba#1.#2.#3{\sandpp#1.#2.#3}
\def\spaa#1.#2.#3.#4{\sandmp#1.{#2#3}.#4}
\def\spbb#1.#2.#3.#4{\sandpm#1.{#2#3}.#4}

\newbox\charbox
\newbox\slabox
\def\s#1{{      
        \setbox\charbox=\hbox{$#1$}
        \setbox\slabox=\hbox{$/$}
        \dimen\charbox=\ht\slabox
        \advance\dimen\charbox by -\dp\slabox
        \advance\dimen\charbox by -\ht\charbox
        \advance\dimen\charbox by \dp\charbox
        \divide\dimen\charbox by 2
        \raise-\dimen\charbox\hbox to \wd\charbox{\hss/\hss}
        \llap{$#1$}
}}

\def\eqn#1{eq.~(\ref{#1})}

\def\e{\epsilon}

\def\Gr{{\rm Gr}}

\def\sign{{\mathop{\rm sign}\nolimits}}

\def\Split{\mathop{\rm Split}\nolimits}
\def\oneloop{{1 \mbox{-} \rm loop}}

\def\stree{{(0)}}
\def\soneloop{{(1)}}

\def\sandp#1.#2.#3{%
\left\langle\smash{#1}{\vphantom1}^{+}\right|{#2}%
\left|\smash{#3}{\vphantom1}^{+}\right\rangle}
\def\ksl{\s{k}}
\def\Ksl{\s{K}}

\def\Soft{{\cal S}}

\def\Res{\mathop{\rm Res}}
\def\tlambda{{\tilde\lambda}}
\def\psl{\s{p}}

\def\Ll{\mathop{\rm L}\nolimits}

\newbox\ourfigbox
\def\SizedFigureWithCaption#1#2#3{%
\setbox\ourfigbox
  \hbox{\hss\epsfxsize #1 \epsfbox{#2}\hss}
\hbox to \wd\ourfigbox{\vbox{\noindent\copy\ourfigbox\break
\vskip -6mm      \hbox to \wd\ourfigbox{\hss#3\hss}}}
}

\def\spa#1.#2{\left\langle#1\,#2\right\rangle}
\def\spb#1.#2{\left[#1\,#2\right]}
\def\spash#1.#2{\spa{\smash{#1}}.{\smash{#2}}}
\def\spbsh#1.#2{\spb{\smash{#1}}.{\smash{#2}}}
\def\lor#1.#2{\left(#1\,#2\right)}
\def\sand#1.#2.#3{%
\left\langle\smash{#1}{\vphantom1}^{-}\right|{#2}%
\left|\smash{#3}{\vphantom1}^{-}\right\rangle}
\def\sandpp#1.#2.#3{%
\left\langle\smash{#1}{\vphantom1}^{+}\right|{#2}%
\left|\smash{#3}{\vphantom1}^{+}\right\rangle}
\def\sandpm#1.#2.#3{%
\left\langle\smash{#1}{\vphantom1}^{+}\right|{#2}%
\left|\smash{#3}{\vphantom1}^{-}\right\rangle}
\def\sandmp#1.#2.#3{%
\left\langle\smash{#1}{\vphantom1}^{-}\right|{#2}%
\left|\smash{#3}{\vphantom1}^{+}\right\rangle}

\begin{document}
\hfuzz 10 pt


\ifpreprint
\noindent
UCLA/05/TEP/4
\hfill SLAC--PUB--11003
\hfill Saclay/SPhT--T05/015
\hfill hep-th/0501240
\fi

\title{On-Shell Recurrence Relations for One-Loop QCD Amplitudes%
\footnote{Research supported in part by the US Department of 
 Energy under contracts DE--FG03--91ER40662 and DE--AC02--76SF00515}}

\author{Zvi Bern}
\affiliation{ Department of Physics and Astronomy, UCLA\\
\hbox{Los Angeles, CA 90095--1547, USA}
}

\author{Lance J. Dixon} 
\affiliation{ Stanford Linear Accelerator Center \\ 
              Stanford University\\
             Stanford, CA 94309, USA
}

\author{David A. Kosower} 
\affiliation{Service de Physique Th\'eorique\footnote{Laboratory 
   of the {\it Direction des Sciences de la Mati\`ere\/}
   of the {\it Commissariat \`a l'Energie Atomique\/} of France.}, 
   CEA--Saclay\\ 
          F--91191 Gif-sur-Yvette cedex, France
}

\date{January 28, 2005}

\begin{abstract}
We present examples of on-shell recurrence relations for
determining rational functions appearing in one-loop QCD
amplitudes. In particular, we give relations for
one-loop QCD amplitudes with all legs of positive helicity, or with one leg of
negative helicity and the rest of positive helicity. Our recurrence
relations are similar to the tree-level ones described by Britto,
Cachazo, Feng and Witten.  A number of new features arise for loop
amplitudes in non-supersymmetric theories like QCD,
including boundary terms and double poles.  We show
how to eliminate the boundary terms, which would interfere 
with obtaining useful relations.  Using the relations we
give compact explicit expressions for the $n$-gluon amplitudes 
with one negative-helicity gluon, up through $n=7$.
\end{abstract}

\pacs{11.15.Bt, 11.25.Db, 11.25.Tq, 11.55.Bq, 12.38.Bx \hspace{1cm}}

\maketitle



\renewcommand{\thefootnote}{\arabic{footnote}}
\setcounter{footnote}{0}


\section{Introduction}
\label{IntroSection}

Seeking out new physics beyond the standard $SU(3)\times
SU(2)\times U(1)$ model of particle interactions is the goal of the
experimental program at the forthcoming Large Hadron Collider (LHC) at
CERN.  The discovery, and study, of such new physics
will depend on our ability to calculate a wide variety of
processes in the gauge theories making up the Standard Model.  
While computations of tree-level scattering amplitudes are a first 
step, the size and scale-variation of the strong coupling constant 
imply that a basic quantitative understanding must also include 
the one-loop amplitudes which enter into next-to-leading order corrections
to cross sections.  Such corrections are required to build a 
theoretical base for a program of precision measurements at 
hadron colliders~\cite{GloverReviewADMP}.   Precision measurements
at the SLAC Linear Collider (SLC) and CERN's Large Electron Positron
(LEP) collider have proven the power of such a program in advancing
our understanding of short-distance physics.

Recently, there has been tremendous progress in understanding and
computing gauge theory scattering amplitudes, stimulated by Witten's
proposal~\cite{WittenTopologicalString}
 of a {\it weak--weak\/} coupling duality between $\NeqFour$
supersymmetric gauge theory and the topological open-string $B$ model,
generalizing Nair's earlier description~\cite{Nair} of the simplest
gauge-theory amplitudes.

While there are still open questions about the precise computation of
tree-level amplitudes in the twistor string theory~\cite{RSV}, and
open questions about the structure of the theory at loop
level~\cite{BerkovitsWitten}, the proposal has already led to several
important insights into the structure of gauge-theory amplitudes.  In
particular, it allowed Cachazo, Svr\v{c}ek and Witten
(CSW)~\cite{CSW} to formulate a new set of vertices and rules for
computing tree-level amplitudes in either supersymmetric or
non-supersymmetric gauge theories.  The vertices are off-shell
continuations of the maximally helicity-violating (MHV) amplitudes,
which were written down long ago by Parke and Taylor~\cite{ParkeTaylor}
--- in amplitude form by Berends and Giele~\cite{BGSix} and
by Mangano, Parke and Xu~\cite{MPX} --- and
proven using recurrence relations by Berends and
Giele~\cite{BGRecurrence}.  (For a discussion of the connections
between different tree-level pictures, see refs.~\cite{Gukov,BBK}.)  A
number of papers have applied these techniques to explicit
calculations~\cite{Khoze,NMHVTree,OtherCSW}.
Recently, several authors have taken the first steps in applying
such ideas to mixed QCD--electroweak amplitudes as
well~\cite{Electroweak}.  At loop level, earlier apparent
complications~\cite{CSWII} in the twistor-space structure of
amplitudes were clarified through an explicit linking of CSW vertices
to the unitarity-based method~\cite{BST}, and via the
elucidation~\cite{CSWIII} and computation of the holomorphic
anomaly~\cite{BBKR,CachazoAnomaly}.

Motivated by early puzzles about the twistor-space structure of
one-loop amplitudes, two groups pursued calculations of the
next-to-MHV (NMHV) class of $n$-gluon
amplitudes in the $\NeqFour$ super-Yang-Mills 
theory~\cite{BCF7,NeqFourSevenPoint,BCFII,NeqFourNMHV,BCFcoplanar}, 
primarily using the unitarity-based
method~\cite{Neq4Oneloop,Neq1Oneloop,UnitarityMachinery,TwoloopSplitting}. 
A recent development, enhancing the power of the unitarity method, is
the observation~\cite{BCFII} that
box integral coefficients can be obtained from generalized unitarity
cuts~\cite{ZFourPartons,TwoloopSplitting,NeqFourSevenPoint} by solving the
constraints that multiple intermediate states be on shell. 
(The notion of `generalized unitarity', as applied to amplitudes for
massive particles, can be traced back to ref.~\cite{Eden}.)
As a by-product of these calculations, ref.~\cite{NeqFourSevenPoint}
revealed a simpler representation for the NMHV seven-gluon amplitude
than had been known previously, from either a recurrence-based
computation~\cite{BGK} or the CSW rules~\cite{CSW}. 
The NMHV computation for arbitrary values of $n$~\cite{NeqFourNMHV},
like that for $n=7$, made use of infrared consistency equations, 
which express the
constraint that a one-loop amplitude must have the correct infrared
poles~\cite{UniversalIR} in the dimensional regulator $\e$.  Again
as a by-product, the solutions to the equations revealed a new class of
representations of tree amplitudes.  A particularly nice
one~\cite{RSVNewTree} gave the natural generalization of the simple
representation to the eight-point amplitude, and suggested a new type
of on-shell recurrence relation for tree amplitudes.  Britto, Cachazo
and Feng wrote down~\cite{BCFRecurrence} the corresponding recurrence
relation, which yields compact expressions for NMHV tree
amplitudes~\cite{LuoWen}. 
The same authors and Witten (BCFW) gave a very simple and elegant 
proof~\cite{BCFW} of the relation using special complex continuations 
of the external momenta.

The proof, which we review in \sect{ProofReviewSection}, is actually
quite general, and applies to any rational function of the external
spinors satisfying certain scaling and factorization properties.  The
generality of the proof motivated us to find a loop version of the
recurrence relation.  In this paper, we shall apply it to the finite
one-loop amplitudes in QCD, which are purely rational.  We will also
argue that it can be applied to compute the rational terms in other,
cut-containing, QCD amplitudes.  We shall see that the application 
to general loop amplitudes in a non-supersymmetric theory like QCD
requires that we address new issues, such as double poles in the complex
analytic continuation, which are related to properties of these 
amplitudes in the limit that two momenta become collinear.

Recurrence relations can provide a technique, complementary
to the unitarity-based method, for the complete calculation of 
one-loop QCD amplitudes.  The unitarity method applies most easily to
terms in the amplitudes that have discontinuities.  Computation
of these terms
requires only knowledge of tree amplitudes evaluated in four dimensions.
The unitarity method can also be applied to computation of 
rational terms, because the latter effectively have cuts at higher 
order in $\e$~\cite{UnitarityMachinery,TwoloopSplitting}; 
but doing so requires knowledge of tree 
amplitudes with two legs in $D=4-2\e$ dimensions, which are harder to compute.
Another approach is to build up an ansatz for the rational terms, based
on their collinear and multi-particle factorization properties.
This approach has been successful for amplitudes with sufficiently
simple structure~\cite{AllPlusA,AllPlus}; and in more complex cases, 
with assistance from other, numerical representations of the 
amplitudes~\cite{ZFourPartons}.  However, it involves some guesswork
and so it is not easy to systematize.  An alternative, recurrence-based 
method, along the lines of the present paper, would represent a 
systematic method for constructing rational function terms in amplitudes
directly from their poles.  It thus holds forth the prospect of 
dramatically simplifying such calculations.

This paper is organized as follows.  In \sect{NotationSection}
we define the notation used in this paper.  \Sect{ProofReviewSection}
reviews the BCFW proof.  Then in \sect{AllPlusSection} we construct 
recurrence relations for the all positive-helicity one-loop amplitudes.
\Sect{OneMinusSection} presents such a relation
for the one-loop amplitudes containing a single negative-helicity gluon,
and also displays new compact forms for the cases of $n=6$ and $n=7$,
obtained via the relation.
Finally, in \sect{ConclusionSection} we give our conclusions and 
outlook for the future.


\section{Notation}
\label{NotationSection}

We use the trace-based color 
decomposition~\cite{TreeColor,BGSix,MPX,TreeReview} 
of amplitudes.  For tree-level amplitudes with $n$ external gluons, 
this decomposition is,
\begin{equation}
{\cal A}_n^\tree(\{k_i,h_i,a_i\}) = 
\sum_{\sigma \in S_n/Z_n} \Tr(T^{a_{\sigma(1)}}\cdots T^{a_{\sigma(n)}})\,
A_n^\tree(\sigma(1^{h_1},\ldots,n^{h_n}))\,,
\label{TreeColorDecomposition}
\end{equation}
where $S_n/Z_n$ is the group of non-cyclic permutations on $n$
symbols, and $j^{h_j}$ denotes the $j$-th momentum and helicity $h_j$.
The $T^a$ are fundamental representation SU$(N_c)$ color matrices
normalized so that $\Tr(T^a T^b) = \delta^{ab}$.  The color-ordered
amplitude $A_n^\tree$ is invariant under a cyclic permutation of its
arguments.

We describe the amplitudes using the spinor helicity formalism.
In this formalism amplitudes are expressed in terms of spinor
inner-products,
\begin{equation}
\spa{j}.{l} = \langle j^- | l^+ \rangle = \bar{u}_-(k_j) u_+(k_l)\,, 
\hskip 2 cm
\spb{j}.{l} = \langle j^+ | l^- \rangle = \bar{u}_+(k_j) u_-(k_l)\, ,
\label{spinorproddef}
\end{equation}
where $u_\pm(k)$ is a massless Weyl spinor with momentum $k$ and plus
or minus chirality~\cite{SpinorHelicity,TreeReview}. Our convention
is that all legs are outgoing. The notation used here follows the
standard QCD literature, with $\spb{i}.{j} = \sign(k_i^0 k_j^0)\spa{j}.{i}^*$
so that,
\begin{equation}
\spa{i}.{j} \spb{j}.{i} = 2 k_i \cdot k_j = s_{ij}\,.
\end{equation}
(Note that the square bracket $\spb{i}.{j}$ differs by an overall sign
compared to the notation commonly used in twistor-space
studies~\cite{WittenTopologicalString}.) 

\def\vmu{{\vphantom{\mu}}}
We denote the sums of cyclicly-consecutive external momenta by
\begin{equation}
K^\mu_{i\cdots j} \equiv 
   k_i^\mu + k_{i+1}^\mu + \cdots + k_{j-1}^\mu + k_j^\mu,
\label{KDef}
\end{equation}
where all indices are mod $n$ for an $n$-gluon amplitude.
The invariant mass of this vector is $s_{i\cdots j} = K_{i\cdots j}^2$.
Special cases include the two- and three-particle invariant masses, 
which are denoted by
\begin{equation}
s_{ij} \equiv K_{i,j}^2
\equiv (k_i+k_j)^2 = 2k_i\cdot k_j,
\qquad \quad
s_{ijk} \equiv (k_i+k_j+k_k)^2.
\label{TwoThreeMassInvariants}
\end{equation}
In color-ordered amplitudes only invariants with cyclicly-consecutive
arguments need appear, {\it e.g.}{} $s_{i,i+1}$ and $s_{i,i+1,i+2}$.
We also write, for the sum of massless momenta belonging to a set $A$,
\be
K^\mu_A \equiv \sum_{a_i \in A}
   k_{a_i}^\mu \,.
\label{KDefAlt}
\end{equation}
Spinor strings, such as
\begin{equation}
 \spba{i}.{\Ksl_A}.{j} = \sum_{a\in A} \spb{i}.{a}\spa{a}.j \,, \hskip 2 cm 
 \spab{i}.{\Ksl_A}.{j} = \sum_{a\in A} \spa{i}.{a}\spb{a}.j
  \,,
\label{longerstrings}
\end{equation}
and
\begin{eqnarray}
\spab{i}.{(a+b)}.{j} &=& \spa{i}.{a} \spb{a}.{j} + \spa{i}.{b} \spb{b}.{j} \,,
          \nonumber   \\
\spbb{i}.{(a+b)}.{(c+d)}.{j} &=& 
     \spb{i}.{a} \spab{a}.{(c+d)}.{j} +
     \spb{i}.{b} \spab{b}.{(c+d)}.{j} \,,
\end{eqnarray}
will also make appearances.  

For one-loop amplitudes, the color decomposition is
similar to the tree-level
case~(\ref{TreeColorDecomposition})~\cite{BKColor}.  
When all internal particles transform in
the adjoint representation of SU$(N_c)$, as is the case for 
$\NeqFour$ supersymmetric Yang-Mills theory, we have
\begin{equation}
{\cal A}_n^\oneloop ( \{k_i,h_i,a_i\} ) =
  \sum_{c=1}^{\lfloor{n/2}\rfloor+1}
      \sum_{\sigma \in S_n/S_{n;c}}
     \Gr_{n;c}( \sigma ) \,A_{n;c}(\sigma) \,,
\label{ColorDecomposition}
\end{equation}
where ${\lfloor{x}\rfloor}$ is the largest integer less than or equal to $x$.
The leading color-structure factor
\begin{equation}
\Gr_{n;1}(1) = N_c\ \Tr (T^{a_1}\cdots T^{a_n} ) \,, 
\end{equation}
is $N_c$ times the tree color factor.  The subleading color
structures are given by
\begin{equation}
\Gr_{n;c}(1) = \Tr ( T^{a_1}\cdots T^{a_{c-1}} )\,
\Tr ( T^{a_c}\cdots T^{a_n}).
\end{equation}
$S_n$ is the set of all permutations of $n$ objects,
and $S_{n;c}$ is the subset leaving $\Gr_{n;c}$ invariant.

The one-loop subleading-color partial amplitudes are given by a sum over
permutations of the leading-color ones~\cite{Neq4Oneloop}.
Therefore we need to compute directly only the leading-color 
single-trace partial amplitudes $A_{n;1}(1^{h_1},\ldots,n^{h_n})$. 


\section{Review of the Proof}
\label{ProofReviewSection}

The proof of the BCFW recurrence relation~\cite{BCFW} starts by introducing a
parameter-dependent shift of two of the external massless spinors,
here labeled $j$ and $l$,
in an $n$-point process,
\begin{eqnarray}
&\tlambda_j &\rightarrow \tlambda_j - z\tlambda_l \,, \nonumber\\
&\lambda_l &\rightarrow \lambda_l + z\lambda_j \,,
\label{SpinorShift}
\end{eqnarray}
where $z$ is a complex number.  This induces a shift of the corresponding
momenta,
\begin{eqnarray}
&p_j^\mu &\rightarrow p_j^\mu(z) = p_j^\mu - 
      {z\over2}{\sand{j}.{\gamma^\mu}.{l}},\nonumber\\
&p_l^\mu &\rightarrow p_l^\mu(z) = p_l^\mu + 
      {z\over2}{\sand{j}.{\gamma^\mu}.{l}} \,,
\label{MomentumShift}
\end{eqnarray}
which preserves their masslessness, $p_j^2(z) = 0 = p_l^2(z)$,
as well as overall momentum conservation.
This shift also implies,
\begin{eqnarray}
&\psl_j &\rightarrow 
  \psl_j(z) = \psl_j 
  - z \, 
  \bigl( \, {|l^-\rangle\langle j^-| + |j^+\rangle\langle l^+|} \, \bigr)\,,
\nonumber\\
&\psl_l &\rightarrow 
  \psl_l(z) = \psl_l
  + z \, 
  \bigl( \, {|l^-\rangle\langle j^-| + |j^+\rangle\langle l^+|} \, \bigr) \,.
\label{SlashedMomentumShift}
\end{eqnarray}

Define
\begin{equation}
A(z) = A(p_1,\ldots,p_j(z),p_{j+1},\ldots,p_l(z),\ldots,p_n),
\end{equation}
which is an on-shell amplitude evaluated at a particular set of complex
momenta.  When $A$ is a tree amplitude or finite one-loop
amplitude, $A(z)$ is a rational function of $z$.
The physical amplitude is given by $A(0)$. 

Now consider the following quantity,
\begin{equation}
{1\over 2\pi i} \oint_C {dz\over z}\,A(z) \,,
\end{equation}
where the contour integral is taken around the circle at infinity.
If $A(z)\rightarrow 0$ as $z\rightarrow\infty$, as in the 
tree-level cases
considered by BCFW, then the integral vanishes, and we can evaluate this
expression as follows,
\begin{equation}
0 = A(0) + \sum_{{\rm poles}\ \alpha} \Res_{z=z_\alpha} {A(z)\over z}\,,
\end{equation}
where the sum is taken over the poles at the $z_{\alpha}$ of $A(z)$.  We
can then solve for $A(0)$,
\begin{equation}
A(0) = -\sum_{{\rm poles}\ \alpha} \Res_{z=z_\alpha}  {A(z)\over z}\,.
\end{equation}

If, on the other hand, $A(z)\rightarrow C_\infty$ as $z\rightarrow \infty$,
the contour integral is equal to $C_\infty$, and this result is modified to,
\begin{equation}
A(0) = C_\infty-\sum_{{\rm poles}\ \alpha} \Res_{z=z_\alpha}  {A(z)\over z} \,.
\end{equation}
As explained in ref.~\cite{BCFW}, so long as $A(z)$ only
has simple poles, each residue is given by factorizing the shifted amplitude
on the appropriate pole in momentum invariants, so that at tree level,
\begin{equation}
A(0) = C_\infty + \sum_{r,s,h} 
   A^h_L(z = z_{rs}) { i \over K_{r\cdots s}^2 } A^{-h}_R(z = z_{rs})  \,,
\label{BCFWRepresentation}
\end{equation}
where $h=\pm1$ labels the helicity of the intermediate state.  There
is generically a double sum, labeled by $r,s$, over momentum poles,
with legs $j$ and $l$ always appearing on opposite sides of the pole.
The squared momentum associated with that pole, $K_{r\cdots s}^2$, is
evaluated in the unshifted kinematics; whereas the on-shell amplitudes
$A_L$ and $A_R$ are evaluated in kinematics that have been shifted 
by \eqn{SpinorShift}, with $z=z_{rs}$.
To extend the approach to one loop, 
the sum~(\ref{BCFWRepresentation})
must also be taken over the two ways of assigning the loop to the pair
$(A_L,A_R)$.

The poles in $z$ arise from poles in momentum invariants; because of
the structure of multiparticle factorization, in general only single
poles arise.  The same is almost true of collinear factorization at
one loop; only lone powers of spinor products --- and hence only
single poles --- appear in the splitting amplitudes in all helicity
configurations except $(+{}+{}+)$ and $(-{}-{}-)$.  In these
identical-helicity cases, the tree splitting amplitude vanishes;
however, the one-loop splitting amplitude has the form
$\spb{a}.{b}/\spa{a}.{b}^2$ or its conjugate~\cite{Neq4Oneloop}, and
double poles do arise.  In no case does anything worse than a double
pole arise.  As we shall see in \sect{OneMinusSection}, the double
poles alter the form of the recurrence relation.

At tree level, it is always possible to find choices of reference momenta
$j$ and $l$ such that there is no boundary term; and only single poles arise.
For the finite one-loop amplitudes we will consider explicitly in this paper,
life is not quite so simple, and
we must confront each of these problems in turn.

\section{All-Positive Helicity One-Loop Amplitudes}
\label{AllPlusSection}

Let us begin with the all-plus amplitude, for which 
Chalmers and the authors~\cite{AllPlusA,AllPlus} wrote a 
collinear-based conjecture~\footnote{%
Note that a version of the `odd' terms $O_n$ in the first reference
in ref.~\cite{AllPlus} (the last line of eq.~(7)) has the wrong sign.},
\begin{equation}
A_{n;1} = {i N_p\over 96\pi^2} 
{H_n\over \spa1.2 \spa2.3\cdots \spa{(n-1)}.n\spa{n}.1},
\label{OneLoopAllPlusAmplitude}
\end{equation}
where
\begin{equation}
H_n = -\sum_{1\leq i_1<i_2<i_3<i_4\leq n} 
  \Tr_{-}\Bigl[\ksl_{i_1}\ksl_{i_2}\ksl_{i_3}\ksl_{i_4}\Bigr] \,,
\end{equation}
and 
\begin{eqnarray}
\Tr_-\Bigl[\ksl_{i_1}\ksl_{i_2}\ksl_{i_3}\ksl_{i_4}\Bigr] &=& {1\over 2}
\Tr[(1-\gamma_5)\ksl_{i_1}\ksl_{i_2}\ksl_{i_3}\ksl_{i_4}] \nonumber\\
&=& \spa{i_1}.{i_2} \spb{i_2}.{i_3} \spa{i_3}.{i_4} \spb{i_4}.{i_1} \,.
\end{eqnarray}
The conjecture was later proven by Mahlon~\cite{Mahlon}.  The factor $N_p$
counts the number of states circulating in the loop, $+1$ or $-1$ for
each adjoint representation bosonic or fermionic state,
respectively.  For Dirac quarks in the fundamental representation, 
the contribution to $N_p$ is $-2/N_c$, so that for QCD 
$N_p = 2(1 -\ n_{\!f}/N_c)$,
where $n_{\! f}$ is the number of quark flavors.
Because the corresponding tree-level amplitude vanishes for this
helicity configuration, the one-loop amplitude
(\ref{OneLoopAllPlusAmplitude}) is finite, and is purely a rational
function of the spinor products.

With a simple shift as given in \eqn{SpinorShift}, it is easy to see
that there must necessarily be a boundary term.  Without loss of generality,
we may take $l = n$.  There are two cases to
consider: when $j$ and $n$ are adjacent, and when they are not.  Take the
second case.  Just as in the tree-level MHV case, two denominator factors
are shifted,
\begin{equation}
\spa{(n-1)}.n \rightarrow \spa{(n-1)}.n +z \spa{(n-1)}.j \,,
\qquad \spa{n}.1 \rightarrow \spa{n}.1 + z \spa{j}.1 \,.
\end{equation}
This gives rise to two distinct single poles in $z$, a result that
could have been obtained without knowledge of the answer, because the
form of the denominator is fixed by general collinear- and 
multiparticle-factorization arguments.

Most terms in $H_n$ are either unshifted, or shifted by only a single power
of $z$.  These terms will fade away as $z\rightarrow \infty$.  The only
dangerous terms are those involving both $\ksl_j$ and $\ksl_n$.  However,
\begin{equation}
{d\over dz}\ksl_j\ksl_n = 0 = {d\over dz}\ksl_n\ksl_j,
\end{equation}
so the terms proportional to $z^2$ are those with 
$i_2=j$ and $i_4=n$,
\begin{eqnarray}
&H_n^\infty &= z^2\sand{j}.{\Ksl_{1\cdots(j-1)}}.n
                \sand{j}.{\Ksl_{(j+1)\cdots (n-1)}}.n\nonumber\\
& &= -z^2\sand{j}.{\Ksl_{1\cdots(j-1)}}.n^2\\
& &= -z^2\sand{j}.{\Ksl_{(j+1)\cdots (n-1)}}.n^2 \,.\nonumber
\end{eqnarray}
Both the numerator and denominator scale as $z^2$, giving rise to a
boundary term.  In the other case, when $j$ and $n$ are adjacent (that
is, $j=1$ or $n-1$), only one pole (one power of $z$) is generated
from the denominator; but correspondingly, the numerator also scales
as $z$, so that again we obtain a boundary term (whose form is somewhat
more complicated, though).

It is not clear how to derive the boundary term without knowing the
complete answer, so any recurrence relation we obtain with one is probably
not useful for {\it deriving\/} the amplitude.  We could, however,
still use it for proving a conjecture. 

Because tree-level amplitudes containing less than two negative-helicity 
gluons vanish, we know that the one-loop identical-helicity 
amplitude has no multi-particle poles, and that the only collinear 
singularities come from the tree-level $(-{};+{}+)$ splitting amplitude.  
This in turn implies that $A(z)$ in this case has only single poles, and
accordingly we can write a recurrence relation for the all-plus
amplitude,
\begin{eqnarray}
A^\soneloop_{n}(1^+,\ldots,n^+) &=& 
 C_\infty +\hskip -10pt
\mathop{\sum_{r=1}^j\sum_{s=j}^{n-1}}_{(r,s)\neq (j,j), (1,n-1)}
\hskip -16pt\sum_{v=0\atop h=\pm}^1
A_{s-r+2}^{(v)}(r^+,\ldots,\hat \jmath^+,\ldots,s^+,-\hat K_{r\cdots s}^h)
{i\over K_{r\cdots s}^2} 
\\ && 
\hskip3cm
  \times
  A_{n-s+r}^{(1-v)}
  (\hat K_{r\cdots s}^{-h},(s+1)^+,\ldots,\hat n^+,\ldots,(r-1)^+)
\nonumber
\end{eqnarray}
(where $j\neq 1,n-1$). For convenience we define $A_n^\stree = A_n^\tree$
and $A_n^\soneloop = A_{n;1}^\oneloop$. The hatted momenta
are evaluated using the shift (\ref{SpinorShift}), with
\begin{equation}
z = z_{rs} =  {K_{r\cdots s}^2 \over \spab{j}.{\Ksl_{r\cdots s}}.{n}} \,.
\end{equation}

Using the observation that tree amplitudes with a lone negative helicity
continue to vanish generically
for $n>3$, even when evaluated at complex momenta~\cite{BCFII},
only four terms survive in the sum,
\begin{eqnarray}
&&{i\over K_{j,j+1}^2} A_3^\stree(\hat \jmath^+,(j+1)^+,-\hat K_{j,j+1}^-)
A_{n-1}^\soneloop(\hat K_{j,j+1}^+,(j+2)^+,\ldots,\hat n^+,\ldots,(j-1)^+)
\nonumber\\
 &&+{i\over K_{j-1,j}^2} A_3^\stree((j-1)^+,\hat \jmath^+,-\hat K_{j-1,j}^-)
A_{n-1}^\soneloop(\hat K_{j-1,j}^+,(j+1)^+,\ldots,\hat n^+,\ldots,(j-2)^+)
\nonumber\\
&&+{i\over K_{n-1,n}^2} 
  A_{n-1}^\soneloop(1^+,\ldots,\hat \jmath^+,\ldots,(n-2)^+,\hat K_{n-1,n}^+)
A_3^\stree(-\hat K_{n-1,n}^-,(n-1)^+,\hat n^+)
\nonumber\\
&&+{i\over K_{n,1}^2} 
A_{n-1}^\soneloop( 2^+,\ldots,\hat\jmath^+,\ldots,(n-1)^+,\hat{K}_{n,1}^+)
A_3^\stree(-\hat K_{n,1}^-,\hat n^+,1^+).
\end{eqnarray}
Here the three-point amplitude, before permuting its arguments 
and shifting its momenta, is given by
\be
A_3^\stree(1^+,2^+,3^-) = - i { {\spb1.2}^3 \over \spb2.3 \spb3.1 } \,.
\label{antiMHV3}
\ee
With the choice of shift we have made, the three-point amplitudes 
$A^\stree(\hat \jmath^+,(j+1)^+,-\hat K_{j,j+1}^-)$ 
and \hbox{$A^\stree((j-1)^+,\hat \jmath^+,-\hat K_{j-1,j}^-)$} 
vanish~\cite{BCFRecurrence}, so that the final relation takes the form,
\begin{eqnarray}
A^\soneloop_{n}(1^+,\ldots,n^+) &=&
 - {i N_p\over 96\pi^2} {\sand{j}.{\Ksl_{(j+1)\cdots (n-1)}}.n^2
                       \over\spa1.2\spa2.3\cdots\spa{(n-2)}.{(n-1)}
                            \spa{(n-1)}.j\spa{j}.1} \nonumber \\
&& \hskip-1.5cm
+{i\over K_{n-1,n}^2}
  A_{n-1}^\soneloop(1^+,\ldots,\hat \jmath^+,\ldots,(n-2)^+,\hat K_{n-1,n}^+)
A_3^\stree(-\hat K_{n-1,n}^-,(n-1)^+,\hat n^+)
\nonumber\\
&& \hskip-1.5cm
+{i\over K_{n,1}^2}
A_{n-1}^\soneloop( 2^+,\ldots,\hat\jmath^+,\ldots,(n-1)^+,\hat{K}_{n,1}^+)
A_3^\stree(-\hat K_{n,1}^-,\hat n^+,1^+),
\end{eqnarray}
where  $z = -\spa{(n-1)}.{n}/\spa{(n-1)}.{j}$
and $z = - \spa{1}.n/\spa1.j$ in the last two lines, respectively.
We have checked this relation numerically through $n=15$, and find that
it is indeed satisfied.

\begin{figure}[t]
\centerline{\epsfxsize 4. truein \epsfbox{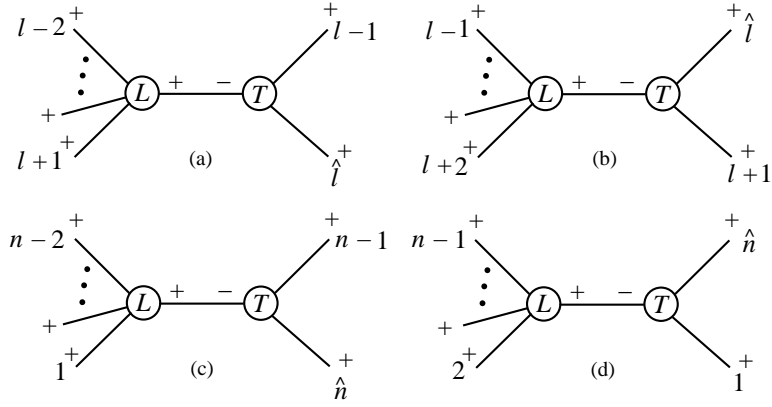}}
\caption{Terms in the recursive expression~(\ref{jlnrecurse}) for 
$A_n^\soneloop(1^+, 2^+,\ldots,n^+)$.  The vertices labeled by a $T$ 
are trees, and the ones labeled by an $L$ are loops.}
\label{AnpFigure}
\end{figure}

Can we find a recurrence which eliminates the boundary
term?  This is indeed possible, so long as we choose a more general 
shift of the original spinors.  In particular, consider the
following shift of the spinors $(j<l<n)$,
\begin{eqnarray}
&\tlambda_j &\rightarrow \tlambda_j - z\tlambda_l 
  - z { \spa{n}.{j} \over \spa{l}.{j} } \tlambda_n,\nonumber\\
&\lambda_l &\rightarrow \lambda_l + z\lambda_j,\label{ThreePointSpinorShift}\\
&\lambda_n &\rightarrow \lambda_n 
   + z { \spa{n}.{j} \over \spa{l}.{j} } \lambda_j,\nonumber
\end{eqnarray}
which also keeps them massless and conserves overall 
four-momentum.  The ratio $\spa{n}.{j}/\spa{l}.{j}$ allows
$z$ to be assigned a definite weight under spinor phase rotations.
So long as we choose $l$ not adjacent to $n$, 
and $j$ adjacent to at most one of $l$ and $n$, 
then we find that $A(z)\rightarrow 0$
as $z\rightarrow\infty$, and we obtain a recurrence free of surface terms.
(We must accordingly take $n\ge 5.$)
The price is that more terms must be included.
For $1 < j < l-1$ and $l < n-1$ (which requires $n\geq6$),
we obtain,
\begin{eqnarray}
&&A^\soneloop_{n}(1^+,\ldots,n^+) = \nonumber\\
&& \hskip0.5cm
{i\over K_{l-1,l}^2}
  A_{n-1}^\soneloop((l+1)^+,\ldots,\hat n^+,1^+,\ldots,\hat \jmath^+,\ldots,
          (l-2)^+,\hat K_{l-1,l}^+)
A_3^\stree(-\hat K_{l-1,l}^-,(l-1)^+,\hat l^+)
\nonumber\\
&& \hskip0.5cm
+{i\over K_{l,l+1}^2}
A_{n-1}^\soneloop
  ((l+2)^+,\ldots,\hat{n}^+, 1^+,\ldots,\hat \jmath^+,\ldots,(l-1)^+,
        \hat K_{l,l+1}^+)
A_3^\stree(-\hat K_{l,l+1}^-,\hat l^+,(l+1)^+) \nonumber\\
&& \hskip0.5cm
+{i\over K_{n-1,n}^2}
  A_{n-1}^\soneloop(1^+,\ldots,\hat \jmath^+,\ldots, \hat{l}^+,\ldots, 
           (n-2)^+,\hat K_{n-1,n}^+)
A_3^\stree(-\hat K_{n-1,n}^-,(n-1)^+,\hat n^+)
\nonumber\\
&& \hskip0.5cm
+{i\over K_{n,1}^2}
A_{n-1}^\soneloop(2^+,\ldots,\hat \jmath^+,\ldots, \hat{l}^+,\ldots, 
   (n-1)^+,\hat K_{n,1}^+)
A_3^\stree(-\hat K_{n,1}^-,\hat n^+,1^+)\,.
\label{jlnrecurse}
\end{eqnarray}
The respective values of the shift variable $z$ in the four 
terms in \eqn{jlnrecurse} are
\be
- { \spa{(l-1)}.{l} \over \spa{(l-1)}.{j} } \,,  \qquad
- { \spa{(l+1)}.{l} \over \spa{(l+1)}.{j} } \,,  \qquad
- { \spa{l}.{j} \over \spa{n}.{j} }
  { \spa{(n-1)}.{n} \over \spa{(n-1)}.{j} }  \,,  \qquad
- { \spa{l}.{j} \over \spa{n}.{j} } { \spa{1}.n \over \spa1.j } \,.
\ee
The recurrence relation~(\ref{jlnrecurse}) is depicted 
diagrammatically in \fig{AnpFigure}.
For $n=5$, there is an analogous relation containing only
three terms.
We have also checked the relation~(\ref{jlnrecurse}) 
numerically through $n=15$.

%

\section{One-Loop Amplitudes with a Single Negative Helicity}
\label{OneMinusSection}

Let us consider next $n$-gluon QCD amplitudes with a single 
negative-helicity gluon, and the remainder of positive helicity.  
These amplitudes were computed by Mahlon for an arbitrary number of
external legs~\cite{Mahlon}, using a loop-level generalization of the
Berends-Giele recurrence relations~\cite{BGRecurrence}.  In this
section, we wish to reproduce these amplitudes, but in a simpler form,
following instead the BCFW recursive methodology.  As we shall see,
the essential difference between the tree amplitudes and the 
one-negative-helicity loop amplitudes is the appearance of double poles.
As a result of their appearance, a straightforward
application of the BCFW recurrence misses single pole terms underlying
the double poles.  We will find a modification which can be written,
for the case at hand, in terms of the soft functions which control 
the universal behavior of amplitudes as
momenta vanish.  In general, boundary terms may arise, but they 
can be avoided by an appropriate choice of a pair of legs $(j,l)$ 
to shift.

It is instructive to first inspect the analytic properties of
the five-point amplitude, as its features will
serve as a guide to finding an all-$n$ recurrence relation.  The one-loop
five-point amplitude with a single negative helicity leg was first calculated
using string-based methods and is given by~\cite{GGGGG,AllPlusA},
\begin{equation}
A_{5}^\soneloop (1^-, 2^+, 3^+, 4^+, 5^+) = i {N_p \over 96 \pi^2} \, 
{1\over \spa3.4^2} 
\Biggl[-{\spb2.5^3 \over \spb1.2 \spb5.1}
       + {\spa1.4^3 \spb4.5 \spa3.5 \over \spa1.2 \spa2.3 \spa4.5^2}
       - {\spa1.3^3 \spb3.2 \spa4.2 \over \spa1.5 \spa5.4 \spa3.2^2} 
     \Biggr] \,. 
\label{mppppsimple}
\end{equation}
Now consider the effect of the shift,
\begin{eqnarray}
&& \lambda_1 \rightarrow \lambda_1, \nonumber \\
&& \tlambda_1 \rightarrow \tlambda_1 - z \tlambda_2\,, \nonumber\\
&& \lambda_2  \rightarrow \lambda_2 + z \lambda_1 \,, \nonumber \\
&& \tlambda_2 \rightarrow \tlambda_2  \,.
\label{SingleMinusShift}
\end{eqnarray}
Applying it to \eqn{mppppsimple} gives the shifted amplitude, 
\begin{eqnarray}
A_{5}^\soneloop (z) &=& i {N_p \over 96 \pi^2} \, 
{1\over \spa3.4^2} 
\Biggl[-{\spb2.5^3 \over \spb1.2 (\spb5.1 - z \spb5.2) }
  + {\spa1.4^3 \spb4.5 \spa3.5 \over \spa1.2  (\spa2.3 + z \spa1.3) \spa4.5^2}
 \nonumber \\
&& \null \hphantom{  i {N_p \over 96 \pi^2} \, {1\over \spa3.4^2} \Biggl[ ]}
       - {\spa1.3^3 \spb3.2 (\spa4.2 + z \spa4.1) \over 
                \spa1.5 \spa5.4 (\spa3.2 + z \spa3.1)^2} 
     \Biggr] \,. 
\label{AzStart}
\end{eqnarray}
Since $A_{5}^\soneloop (z)$ vanishes as $z \rightarrow \infty$, the
choice of shift (\ref{SingleMinusShift}) does not generate a boundary
term at $z\rightarrow \infty$.  Other shifts are also possible,
but not all are as useful; some do generate boundary terms.  Our
shifted amplitude (\ref{AzStart}) has a new feature: the appearance of
a double pole in $z$.  The double pole implies that the BCFW recurrence
relation cannot be applied directly, because it relied on the amplitude having
only single
poles, as is the case at tree level.  Nevertheless, we will use \eqn{AzStart}
to track down the required modifications.

To proceed with our investigation, rewrite \eqn{AzStart} as a sum over
pole terms,
\begin{eqnarray}
A_{5}^\soneloop (z) &=& i {N_p \over 96 \pi^2} \, 
{1\over \spa3.4^2} 
\Biggl[-{\spb2.5^3 \over \spb1.2 (\spb5.1 - z \spb5.2) }
  + {\spa1.4^3 \spb4.5 \spa3.5 \over \spa1.2  (\spa2.3 + z \spa1.3) \spa4.5^2}
 \nonumber \\
&& \null \hskip 3 cm 
   - {\spa1.3^3 \spb3.2 \spa4.3 \spa2.1  \over \spa1.5 \spa5.4 \spa3.1
                  (\spa3.2 + z \spa3.1)^2} \nonumber\\
&& \null \hskip 3 cm 
   - {\spa1.3^3 \spb3.2 \spa4.1 \over \spa1.5 \spa5.4 \spa3.1
                  (\spa3.2 + z \spa3.1)} 
     \Biggr] \,, 
\label{A5z}
\end{eqnarray}
where we have used the Schouten identity to expose the single pole
sitting under the double pole. Taking $z=0$, we obtain a form of
the amplitude whose terms we wish to map onto terms of a recurrence
relation,
\begin{eqnarray}
A_{5}^\soneloop (z=0) &=& i{N_p \over 96 \pi^2} \, 
\Biggl[-{\spb2.5^3 \over \spa3.4^2\spb1.2 \spb5.1 }
  + {\spa1.4^3 \spb4.5 \spa3.5 \over \spa3.4^2\spa1.2  \spa2.3 \spa4.5^2}
 \nonumber \\
&& \null \hskip 1 cm 
   + {\spa1.3^2 \spb3.2 \spa2.1  \over \spa3.4 \spa1.5 \spa4.5
                  \spa3.2^2}
   + {\spa1.3^2 \spb3.2 \spa4.1 \over \spa3.4^2 \spa1.5 \spa5.4 
                  \spa3.2}  \Biggr] \,.  \hskip 1 cm 
\label{mppppTarget}
\end{eqnarray}

The appearance of a double pole in
\eqn{A5z} may seem puzzling at first sight.
It can be understood from the structure of the one-loop
three-vertex with identical helicities, used for obtaining one-loop
splitting amplitudes~\cite{Neq4Oneloop}. When the momenta of legs 1
and 2 become collinear, the vertex
becomes~\cite{BernChalmers,OneloopSplit}
\begin{eqnarray}
A_3^\soneloop(1^+, 2^+, 3^+) & = &
  - i {N_p\over 96 \pi^2} {\pol^{(+)}_3 \cdot (k_1 - k_2)  \over 
               \sqrt{2} } \biggl[\pol_1^{(+)} \cdot \pol_2^{(+)}
              - 2 {k_1 \cdot \pol_2^{(+)} \,
                    k_2 \cdot \pol_1^{(+)} \over K_{12}^2} \biggr]
      \nonumber \\
& = &- i {N_p \over 96 \pi^2}
       {\spb1.2 \spb2.3 \spb3.1  \over K_{12}^2} \,.
\label{OneLoopThreeAmplitude}
\end{eqnarray}
It is useful to expose the kinematic pole, so we define
\begin{equation}
A_{3}^\soneloop(1^+, 2^+, 3^+) \equiv 
          {1 \over K_{12}^2} V_3^\soneloop(1^+, 2^+, 3^+) \,.
\label{OneLoopThreeVertex}
\end{equation}
The propagator of the factorized leg provides a {\it second\/}
factor of $1/K_{12}^2$, leading to the observed double poles in the amplitude.
These poles make their explicit appearance in \eqn{mppppsimple} as
spinor products $\spa{2}.{3}$, $\spa{3}.{4}$ and $\spa{4}.{5}$ appearing
quadratically in the denominator.  (They can also be understood
as arising from those collinear singularities, with $k_a$ parallel to $k_b$,
and $k_a \approx x(k_a+k_b)$, which involve the one-loop
splitting amplitude denoted by~\cite{Neq4Oneloop}
$\Split_{+}^\oneloop(x,a^+,b^+) = 
- i N_p/(96\pi^2) \times \sqrt{x(1-x)} \spb{a}.{b}/{\spa{a}.{b}}^2$.)

We turn now to the recursive construction of the five-point amplitude.  
Our input expressions are the one-loop vertex~(\ref{OneLoopThreeVertex}) 
as well as the one-loop four-point amplitudes~\cite{BKStringBased},
\begin{eqnarray}
A_{4}^\soneloop (1^+,2^+,3^+,4^+)  &=&  - i \, {N_p \over 96 \pi^2} \,
  {\spb1.2 \spb3.4 \over \spa1.2 \spa3.4}\,,
\label{pppp} \\
A_{4}^\soneloop (1^-,2^+,3^+,4^+)  &=&  i \, {N_p \over 96 \pi^2} \,
     {\spa2.4 \spb2.4^3 \over \spb1.2 \spa2.3 \spa3.4 \spb4.1}\,.
\label{mppp}
\end{eqnarray}
Although the symmetry is not manifest in the form we have written it, 
$A_{4}^\soneloop (1^+,2^+,3^+,4^+)$ is invariant
under cyclic permutations of its arguments. 
We shall also need the MHV $n$-gluon tree 
amplitudes~\cite{ParkeTaylor,MPX,BGRecurrence},
\be
A_n^\stree(1^+,\ldots,(j-1)^+,j^-,(j+1)^+,\ldots,
                      (k-1)^+,k^-,(k+1)^+,\ldots,n^+)
 = i { {\spa{j}.{k}}^4 \over \spa1.2 \spa2.3 \cdots \spa{n}.1 } \,,
\label{MHVtree}
\ee
and the three-point amplitude for $({+}{+}{-})$, given in \eqn{antiMHV3}.

%
\begin{figure}[t]
\centerline{\epsfxsize 4. truein \epsfbox{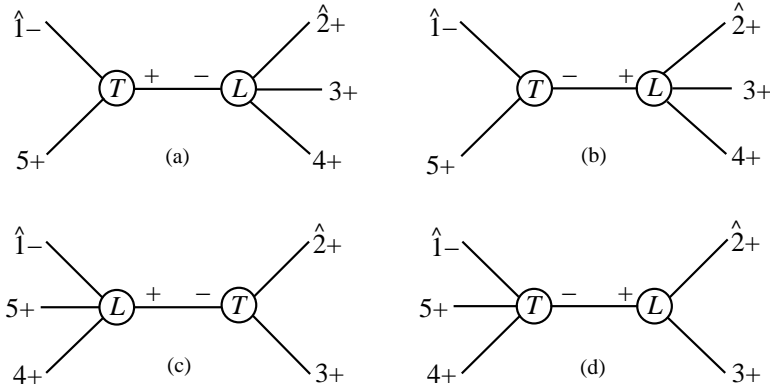}}
\caption{Terms in the recursive expression for 
$A_5^\soneloop(1^-, 2^+, 3^+, 4^+,5^+)$.}
\label{A5mFigure}
\end{figure}

The proof of the recurrence relations in ref.~\cite{BCFW}
leads us to expect a recurrence relation for 
the five-point one-loop amplitude of the form,
\begin{eqnarray}
A_5^\soneloop(1^-, 2^+, 3^+, 4^+, 5^+) &\sim &
 A_{3}^\stree (\hat 1^-, -\hat K_{15}^+, 5^+)  \, {i \over K_{15}^2} \, 
A_{4}^\soneloop(\hat 2^+, 3^+, 4^+, \hat K_{15}^-) \nonumber \\
&&\null \hskip .3 cm 
+ 
A_{3}^\stree(\hat 1^-, -\hat K_{15}^-, 5^+) \, {i \over K_{15}^2} \, 
A_{4}^\soneloop(\hat 2^+, 3^+, 4^+, \hat K_{15}^+) \nonumber \\
&& \null \hskip .3 cm 
+
A_{4}^\soneloop(\hat 1^-, \hat K_{23}^+, 4^+,5^+) \, {i \over K_{23}^2} \, 
A_{3}^\stree(\hat 2^+, 3^+,-\hat K_{23}^-)  \nonumber \\
&& \null \hskip .3 cm 
+
A_{4}^\stree(\hat 1^-, \hat K_{23}^-, 4^+,5^+) \, {i \over (K_{23}^2)^2} \, 
V_{3}^\soneloop(\hat 2^+, 3^+,-\hat K_{23}^+) \,,
\label{FivePtRecursion}
 \end{eqnarray}
corresponding to the four diagrams illustrated in \fig{A5mFigure}.
Following the discussion of \sect{ProofReviewSection}, it is 
straightforward to evaluate each term.  As we shall see, however, these
terms do not yield the complete answer,
because they miss a single pole
lying underneath the double pole of the last term.
However, an inspection of the discrepancy will allow us to determine an 
$n$-point recurrence relation.

The first diagram, shown in \fig{A5mFigure}(a), corresponding to the 
first term in \eqn{FivePtRecursion}, simply vanishes,
\begin{equation}
D^{(a)} = 
A_{3}^\stree(\hat 1^-, -\hat K_{15}^+, 5^+) \, {i \over K_{15}^2} \, 
A_{4}^\soneloop(\hat 2^+, 3^+, 4^+, \hat K_{15}^-) = 0 \,,
\end{equation}
because~\cite{BCFRecurrence}
\begin{equation}
A_3^\stree (\hat 1^-, -\hat K_{15}^+, 5^+) 
\propto \spbsh{(-\hat K_{15})}.5^3 
\propto \sand1.{(1+5)}.5^3 = 0\,.
\end{equation}

\def\vK{\vphantom{{\hat K}}}
We may evaluate the next diagram, shown in \fig{A5mFigure}(b), 
corresponding to the second term in \eqn{FivePtRecursion}, to obtain,
\begin{eqnarray}
D^{(b)}& =& A_{3}^\stree(\hat 1^-, -\hat K_{15}^-, 5^+) \,
            {i \over K_{15}^2} \, 
A_{4}^\soneloop(\hat 2^+, 3^+, 4^+, \hat K_{15}^+) \nonumber \\
   & = &   i\, {N_p \over 96 \pi^2} {\spash{\hat 1}.{(-\hat K_{15})}^3 \over 
              \spash{\hat 1}.5 \spash{5}.{(-\hat K_{15})}\vK }
              \times {1 \over K_{15}^2}
              \times {\spb3.4 \spbsh{\hat 2}.{\hat K_{15}} \over
                      \spa3.4 \spash{\hat 2}.{\hat K_{15}} \vK} \nonumber \\
   & = & 
i \, {N_p \over 96 \pi^2} {\sand1.{K_{15}}.2^3 \over 
             \omega {\overline\omega} \spa1.5\sand{5}.{K_{15}}.2 }
              \times {1 \over K_{15}^2}
              \times {\spb3.4 \sand{1}.{K_{15}}.2 \over
                      \spa3.4 
                      \sand{\hat 2}.{K_{15}}.2} \,,
\label{Diagramb}
\end{eqnarray}
where $\omega = \spbsh{\hat K_{15}}.{2}$ and 
$\bar\omega = \spash{1}.{\hat K_{15}}$.
Using
\begin{equation}
 \sand{\hat 2}.{K_{15}}.2 = \sand{2}.{(1+5)}.2 +
   {K_{15}^2 \over \sand1.{(1+5)}.2} \, \sand1.{(1+5)}.2 = K_{34}^2 \,,
\end{equation}
and
\begin{equation}
 \omega {\overline\omega} = \sand1.{K_{15}}.2 = \spa1.5 \spb5.2 \,,
\end{equation}
diagram (b) of \fig{A5mFigure} simplifies to 
\begin{equation}
D^{(b)}  = 
  - i {N_p \over 96 \pi^2} {\spb2.5^3 \over \spa3.4^2 \spb1.2 \spb5.1} \,.
\end{equation}
This matches precisely the first term in our target expression
(\ref{mppppTarget}). 

Similarly, it is not difficult to work out the diagram shown in 
\fig{A5mFigure}(c), with 
the result
\begin{eqnarray}
D^{(c)} &=& 
A_{4}^\stree(\hat 1^-, \hat K_{23}^+, 4^+,5^+) \, {i \over K_{23}^2} \, 
A_{3}^\stree(\hat 2^+, 3^+, -\hat K_{23}^-)  \nonumber \\
&= & i {N_p \over 96 \pi^2} 
 {\spa1.4^3 \spb4.5 \spa3.5 \over \spa3.4^2\spa1.2  \spa2.3 \spa4.5^2} \,,
\end{eqnarray}
matching the second term of \eqn{mppppTarget}.  Finally, 
for diagram (d) we obtain the double-pole contribution very easily, 
\begin{eqnarray}
D_2^{(d)} &=& 
A_{4}^\stree(\hat 1^-, \hat K_{23}^-, 4^+,5^+) \, {i \over (K_{23}^2)^2} \, 
V_{3}^\soneloop(\hat 2^+, 3^+, -\hat K_{23}^+) \nonumber \\
&=& i\, {N_p \over 96 \pi^2} 
{\spa1.3^2 \spb3.2 \spa2.1  \over \spa3.4 \spa1.5 \spa4.5 
                  \spa3.2^2} \,,
\label{DoublePoleDiagram}
\end{eqnarray}
matching the third term of the target expression (\ref{mppppTarget}).

We now must confront the question of constructing the single-pole
contribution in diagram (d) which underlies the double pole.
This contribution must produce the fourth term in~\eqn{mppppTarget}.
Notice that this term contains a factor of $\spb3.2/\spa3.2$.
Thus it should be dictated by the behavior of 
$A_5^\soneloop(1^-, 2^+, 3^+, 4^+, 5^+)$ as the complex momenta
$k_2$ and $k_3$ become parallel.  While for complex momenta
the factor $\spb3.2/\spa3.2$ can be singular as $k_2\cdot k_3 \to 0$,
for real momenta it is always nonsingular, because 
$|\spa2.3| = |\spb2.3| = \sqrt{|2 k_2\cdot k_3|}$.
(Even for real momenta, the ratio does depend on an azimuthal 
rotation angle in this limit, in which $k_2$ and $k_3$ 
are rotated while holding $k_2+k_3 $ and $k_2\cdot k_3$ fixed.)
As $k_2\cdot k_3 \to 0$, the factor $\spb3.2/\spa3.2$ is subleading
with respect to the leading collinear behavior captured by the 
one-loop splitting amplitudes,
$\Split_{+}^\oneloop(x,2^+,3^+) \propto \spb{2}.{3}/{\spa{2}.{3}}^2$
and $\Split_{-}^\oneloop(x,2^+,3^+) \propto 1/{\spa{2}.{3}}$.
The usual discussions of universal collinear behavior of amplitudes 
limits~\cite{Neq4Oneloop,BernChalmers} 
do not extend to this level of accuracy.

Instead we proceed by experimentation.
To motivate a guess, we assume the single-pole term is related
to the double-pole term~(\ref{DoublePoleDiagram})
associated with $\hat{K}_{23}^2 = 0$.  Hence we try an ansatz
which is equal to this expression, multiplied by some function which
is suppressed as $k_2\cdot k_3 \to 0$.   The additional
multiplicative function should be dimensionless, and 
invariant under phase rotations of spinors associated with all 
external states, and the intermediate state $\hat{K}_{23}$.
We can use universal multiplicative ``soft factors,''  which 
describe the insertion
of a soft gluon $s$ between two hard partons $a$ and $b$
in a color-ordered amplitude, to construct such a multiplicative
function.  The soft factors depend only on the helicity of the 
soft gluon and are given by~\cite{TreeReview},
\begin{eqnarray}
\Soft^\stree( a,  s^+, b) & =& {\spa{a}.{b} \over \spa{a}.{s} \spa{s}.{b}} \,,\\
\Soft^\stree( a,  s^-, b) & =& -{\spb{a}.{b} \over \spb{a}.{s} \spb{s}.{b}} \,.
\label{softdef}
\end{eqnarray}
They are invariant under phase rotations of spinors associated with 
$a$ and $b$, but not $s$.  However, the product
\begin{equation}
K_{23}^2 \Soft^\stree(a,  s^+, b) \Soft^\stree(c,  (-s)^-, d) 
\label{trysoft}
\end{equation}
is dimensionless, invariant under phase rotations of
$a$, $b$, $c$, $d$ and $s$, and suppressed as $K_{23}^2 \to 0$
(for suitable choices of $a$, $b$, $c$, $d$ and $s$).

Since $s$ appears in two factors in \eqn{trysoft} carrying
opposite helicity, it is natural to identify it with the on-shell 
intermediate momentum $\hat{K}_{23}$.  Choosing $c=3$ and $d=\hat{2}$
produces the desired collinear behavior $\propto \spa2.3$
for~\eqn{trysoft}.
With a little more experimentation, we find that in this case, the
color-adjacent legs $\hat{1}$ and $4$ should
be identified with $a$ and $b$.  Thus we arrive
at the candidate single-pole term given by
\begin{equation}
D_1^{(d)} = 
 A_{4}^\stree(\hat 1^-, \hat K_{23}^-, 4^+,5^+) \,
\Soft^\stree(\hat 1,  \hat K_{23}^+, 4)  {i \over K_{23}^2} \, 
\Soft^\stree(3,  -\hat K_{23}^-, \hat 2) \,
V_{3}^\soneloop(\hat 2^+, 3^+, -\hat K_{23}^+) \,. \hskip .5 cm 
\label{SinglePoleAnsatz}
\end{equation}
Evaluating \eqn{SinglePoleAnsatz} we have
\begin{equation}
D_1^{(d)} =  i {N_p \over 96 \pi^2} 
  {\spa1.3^2 \spb3.2 \spa4.1 \over \spa3.4^2 
         \spa1.5 \spa5.4 \spa3.2} \,,
\end{equation}
so it reproduces the last term in the amplitude~(\ref{mppppTarget}).

We now take the value of the diagram in \fig{A5mFigure}(d) to be the
sum of the single- and double-pole contributions,
\begin{eqnarray}
D^{(d)} &= & D_1^{(d)} + D_2^{(d)} \nonumber \\
& = & i {N_p \over 96 \pi^2} \Biggl[
 {\spa1.3^2 \spb3.2 \spa2.1  \over \spa3.4 \spa1.5 \spa4.5 
                  \spa3.2^2}
 + {\spa1.3^2 \spb3.2 \spa4.1 \over \spa3.4^2 
         \spa1.5 \spa5.4 \spa3.2} \Biggr] \,, 
\end{eqnarray}
which matches the last two terms in the amplitude~(\ref{mppppTarget}). 

We have thus succeeded in finding a simple interpretation of
the five-point amplitude (\ref{mppppTarget}) in terms of a recursive 
diagrammatic construction,
\begin{equation}
A_5^\soneloop(1^-, 2^+, 3^+, 4^+, 5^+)
 =  D^{(a)} + D^{(b)} + D^{(c)} + D^{(d)}\,,
\end{equation}
corresponding to the diagrams in \fig{A5mFigure}, 
after interpreting diagram (d) as a sum over single-pole and double-pole
contributions.

%
\begin{figure}[t]
\centerline{\epsfxsize 4. truein \epsfbox{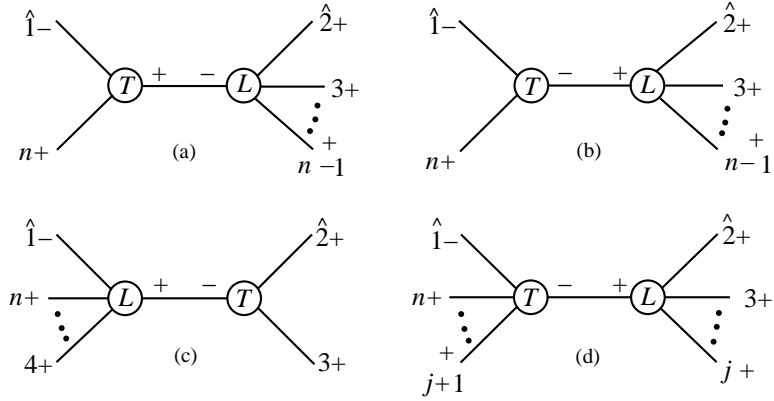}}
\caption{The diagrams describing a recursive formula for 
$A_n^\soneloop(1^-, 2^+, \ldots, n^+)$.  Diagram (a) vanishes, while
diagram (d) needs to be summed over $j$.  For $j=3$ diagram (d) 
contains both single-pole and double-pole contributions.
 }
\label{AnmFigure}
\end{figure}

To generalize this construction to $n$ legs, we assume that the term
corresponding to the single pole underlying the double pole in
$K_{23}^2$ is given by the double-pole contribution, multiplied by the
same function found to work for $n=5$.  The resulting recursion
relation, shown diagrammatically in \fig{AnmFigure}, is
\begin{eqnarray}
&& A^\soneloop_n(1^-, 2^+, \ldots, n^+) \nonumber\\
&& \null \hskip 1 cm 
 = 
A_{n-1}^\soneloop(4^+, 5^+,  \ldots, n^+, \hat 1^-, \hat K_{23}^+)  \,
   {i \over K_{23}^2} \, A_{3}^\stree(\hat 2^+, 3^+,-\hat K_{23}^-)  
 \nonumber \\
&& \null  \hskip 1 cm 
+
\sum_{j=4}^{n-1}
 A_{n-j+2}^\stree((j+1)^+,(j+2)^+, \ldots, n^+, \hat 1^-, \hat K_{2\ldots j}^-) \, 
 {i \over K_{2\ldots j}^2} \, 
  A_{j}^\soneloop(\hat 2^+, 3^+, \ldots, j^+, -\hat K_{2\ldots j}^+)
 \nonumber \\
&& \null \hskip 1 cm 
+
A_{n-1}^\stree(4^+,5^+, \ldots, n^+, \hat 1^-, \hat K_{23}^-) \, 
 {i \over (K_{23}^2)^2} \, 
 V_{3}^\soneloop(\hat 2^+, 3^+, -\hat K_{23}^+) \nonumber \\
&& \null  \hskip 2.5cm 
 \times \biggl( 1 + K_{23}^2 \, 
               \Soft^\stree(\hat 1,   \hat K_{23}^+, 4) \,
               \Soft^\stree(3,  -\hat K_{23}^-, \hat 2) \biggr) \,.
\label{NptRecursion}
\end{eqnarray}
The hatted momenta are evaluated using the shift~(\ref{SingleMinusShift}),
where the value of $z$ 
for the $K_{2\ldots j}$ channel is,
\begin{equation}
z = - { K_{2\ldots j}^2 \over \spab{1}.{\Ksl_{2\ldots j}}.{2} }\,.
\end{equation}

For self-consistency, amplitudes generated by this
recurrence relation must vanish under the shift (\ref{SingleMinusShift}) in
the limit $z \rightarrow \infty$.  Otherwise, there would be a
boundary term not included in the recurrence relation.  To show that the
amplitudes generated by \eqn{NptRecursion} satisfy this criterion, we
first observe that under the shift the explicit poles appearing in the
amplitude will behave either as $1/z$ for the single poles or $1/z^2$
for the $1/(K_{23}^2)^2$ double pole.   In fact, these are the {\it only}
places $z$-dependence will arise.  (Basically, the point of the 
recursive decomposition is to expose the $z$-dependence as 
pole factors multiplied by $z$-independent residues.)
To see this explicitly, first note that $\lambda_1$ is 
unaffected by the shift, while $\tlambda_1$ is affected.
In the $j$th term in \eqn{NptRecursion},
consider the spinor product $\spb{1}.{a}$ for an arbitrary
leg $a$ (not equal to $\hat K^-_{2\cdots j}$).
It looks as though it might develop $z$-dependence in the next shift.
However, what actually appears is the hatted expression,
\bea
\spb{\hat 1}.{a} &=& \spb{1}.{a}
 + { K_{2\cdots j}^2 \over \spab1.{\Ksl_{2\cdots j}}.{2} } \spb{2}.{a}
= { \spbb{2}.{\Ksl_{(j+1)\cdots 1}}.{\ksl_1}.{a}
  - \spbb{2}.{\Ksl_{(j+1)\cdots 1}}.{\Ksl_{(j+1)\cdots 1}}.{a} 
 \over \spab1.{\Ksl_{(j+1)\cdots 1}}.{2} }
\nonumber \\
&=& - { \spbb{2}.{\Ksl_{(j+1)\cdots 1}}.{\Ksl_{(j+1)\cdots n}}.{a}
   \over \spab1.{\Ksl_{(j+1)\cdots n}}.{2} } \,.
\label{shift1a}
\eea
Since $\lambda_1$ and $\tlambda_2$ are unaffected by the shift,
the only possible source of $z$-dependence is from the $\ksl_1$ inside
$\Ksl_{2\cdots j}$.  But this term is proportional to $\spb2.1$,
and thus unaffected (because the $\tlambda_1$ shift is proportional
to $\tlambda_2$).   

Similarly, the expressions
on the other side of the $K_{2\cdots j}^2$ pole which might
appear to develop $z$-dependence are of the form
\bea
\spa{\hat 2}.{b} &=& \spa{2}.{b}
 - { K_{2\cdots j}^2 \over \spab1.{\Ksl_{2\cdots j}}.{2} } \spa{1}.{b}
\nonumber \\
&=& - { \spaa{1}.{\Ksl_{2\cdots j}}.{\Ksl_{3\cdots j}}.{b}
   \over \spab1.{\Ksl_{2\cdots j}}.{2} } \,,
\label{shift2b}
\eea
and so they also are easily seen to be unaffected by the 
subsequent $z$ shift.  An expression involving the intermediate
hatted momentum, like $\spash{\hat 2}.{\hat K_{2\cdots j}}$,
can be re-written as 
\bea
\spash{\hat 2}.{\hat K_{2\cdots j}}
= { \spab{\hat 2}.{\Ksl_{2\cdots j}}.{2}
  \over \spab{1}.{\Ksl_{2\cdots j}}.{2} } \,,
\eea
and then $\Ksl_{2\cdots j}$ can be split into a sum of momenta
in the numerator, applying \eqn{shift2b} to each term.  The denominator
is also inert.  Thus the only $z$-dependence in \eqn{NptRecursion}
is from the  explicit $K_{2\cdots j}^2$ pole factors, and
the expression vanishes as $z \to \infty$.

We have checked the recurrence relation (\ref{NptRecursion}) 
numerically against the computation of Mahlon~\cite{Mahlon} through
$n=15$.  The relation leads straightforwardly to compact expressions 
for higher-point amplitudes. For example, the six-point amplitude can be
written as,
\begin{eqnarray}
&&A_{6;1}(1^-,2^+,3^+,4^+,5^+,6^+)\nonumber \\ 
&& \hskip0.5cm = i \, {N_p \over 96\pi^2} \Biggl[
 { {\spab1.{(2+3)}.6}^3
    \over \spa1.2 \spa2.3 {\spa4.5}^2 \, s_{123} \, \spab3.{(1+2)}.6 }
+ { {\spab1.{(3+4)}.2}^3
    \over {\spa3.4}^2 \spa5.6 \spa6.1 \, s_{234} \, \spab5.{(3+4)}.2 }
\nonumber \\ 
&& \hskip2.0cm
+ { {\spb2.6}^3 \over \spb1.2 \spb6.1 \, s_{345} } \Biggl(
      { \spb2.3 \spb3.4 \over \spa4.5 \, \spab5.{(3+4)}.2 }
    - { \spb4.5 \spb5.6 \over \spa3.4 \, \spab3.{(1+2)}.6 }
    + { \spb3.5 \over \spa3.4 \spa4.5 } \Biggr)
\nonumber \\ 
&& \hskip2.0cm
- { {\spa1.3}^3 \spb2.3 \spa2.4
     \over {\spa2.3}^2 {\spa3.4}^2 \spa4.5 \spa5.6 \spa6.1 }
+ { {\spa1.5}^3 \spa4.6 \spb5.6
     \over \spa1.2 \spa2.3 \spa3.4 {\spa4.5}^2 {\spa5.6}^2 }
\nonumber \\ 
&& \hskip2.0cm
- { {\spa1.4}^3 \spa3.5 \spab1.{(2+3)}.4
     \over \spa1.2 \spa2.3 {\spa3.4}^2 {\spa4.5}^2 \spa5.6 \spa6.1 } 
\Biggr] \,.
\label{mpppppsimple}
\end{eqnarray}
Similarly, in the seven-point case we obtain the form,
\begin{equation}
A_{7;1}(1^-,2^+,3^+,4^+,5^+,6^+,7^+) = 
\hat{A}_7^- + \hat{A}_7^- \bigr|_{\rm flip} \,,
\label{mppppppREF}
\end{equation}
where
\begin{equation}
X(1,2,3,4,5,6,7) \bigr|_{\rm flip} = - X(1,7,6,5,4,3,2) \,,
\label{mppppppflipdef}
\end{equation}
and
\begin{eqnarray}
\hat{A}_7^- &=& i \, {N_p \over 96\pi^2} \Biggl[
 { {\spab1.{(2+3)}.7}^3 
   \over \spa1.2 \spa2.3 \spa4.5 \spa5.6 \, s_{123} \, s_{456} 
    \, \spab3.{(1+2)}.7 } 
\nonumber \\ 
&& \hskip2.5cm
\times
  \Biggl( { s_{45} \spab1.{(2+3)}.4 \over \spaa1.{(2+3)}.{(4+5)}.6 }
    + { \spbb6.{(4+5)}.{(5+6)}.7 \over \spab4.{(5+6)}.7 } \Biggr)
\nonumber \\ 
&& \hskip1.1cm
 + { {\spab1.{(5+6)}.7}^3 
    \over \spa1.2 \spa2.3 \spa3.4 {\spa5.6}^2 \, s_{567} 
     \, \spab4.{(5+6)}.7 }
\nonumber \\ 
&& \hskip1.1cm
+ { {\spb2.7}^3 
    \over \spb1.2 \spa3.4 \spa4.5 \spa5.6 \spb7.1 \, s_{712} }
  \Biggl( {1\over2} \spb3.6
   + { \spbb6.{(4+5)}.{(5+6)}.7 \over \spab3.{(1+2)}.7 }
\nonumber \\ 
&& \hskip6.5cm
   - {1\over2} { s_{45} \,  \spbb2.{(3+4)}.{(5+6)}.7 
         \over \spab3.{(1+2)}.7 \, \spab6.{(7+1)}.2 } \Biggr)
\nonumber \\ 
&& \hskip1.1cm
- { {\spa1.3}^3 \spb2.3 \spa2.4
    \over {\spa2.3}^2 {\spa3.4}^2 \spa4.5 \spa5.6 \spa6.7 \spa7.1 }
- { {\spa1.4}^3 \spa3.5 \spab1.{(2+3)}.4
    \over \spa1.2 \spa2.3 {\spa3.4}^2 {\spa4.5}^2 \spa5.6 \spa6.7 \spa7.1 }
\nonumber \\ 
&& \hskip1.1cm
- {1\over2} { {\spaa1.{(2+3)}.{(4+5)}.1}^3 
    \over \spa1.2 \spa2.3 {\spa4.5}^2 \spa6.7 \spa7.1
    \, \spaa1.{(6+7)}.{(4+5)}.3 \, \spaa1.{(2+3)}.{(4+5)}.6 }
\Biggr] \,.
\nonumber \\
\label{mppppppsimple}
\eea

Although the amplitudes~(\ref{mpppppsimple}) and (\ref{mppppppsimple}) 
are quite compact in comparison with the expressions appearing in
ref.~\cite{Mahlon}, they do contain spurious denominators of the 
form $\spab{a}.{(b+c)}.d$ and $\spaa{a}.{(b+c)}.{(d+e)}.f$.  
Such denominators typically arise
from reducing pentagon integrals to lower-point integrals.
It is of course no surprise that we generate such denominators, 
given the origin of the BCFW-type recurrences in the relations between
coefficients of one-loop amplitudes.
Similar spurious denominators also arise in the CSW construction 
of amplitudes from MHV vertices.  In fact, it appears empirically that
spurious denominators are {\it necessary} to obtain the most 
compact forms of amplitudes that contain many multi-particle factorization
poles.

\section{Conclusions and Outlook}
\label{ConclusionSection}

In this paper, we presented a recursive approach to obtaining rational
terms appearing in non-supersymmetric one-loop amplitudes, based on
the Britto, Cachazo, Feng and Witten~\cite{BCFRecurrence,BCFW}
recursive approach to tree-level amplitudes.  A striking feature of
the proof of the method in ref.~\cite{BCFW} is that it relies only on
very general properties of field theory and complex function
theory. It should therefore apply much more broadly than to massless
gauge theory tree amplitudes.  This motivated us to revisit the
question of loop amplitudes.

In this paper, as an initial application, we focused on the finite
one-loop all-gluon QCD amplitudes which are purely rational: 
amplitudes with all gluons of positive helicity, and those with a 
lone negative-helicity gluon.
These amplitudes had been obtained previously in
refs.~\cite{AllPlusA,AllPlus,Mahlon}, and are therefore quite useful as a
laboratory for studying the new features appearing at loop level.  Two
essential differences arising at loop level are boundary terms
and double poles, both of which modify the form of the recurrence.  The
boundary terms interfere with obtaining a useful recurrence relation, 
but via an appropriate choice of shifts we can avoid these.  
From general field theory
considerations, tree-level amplitudes cannot have double poles in
kinematic invariants.  However, at loop level double poles do arise,
when using complex momenta.  These double poles induce extra terms in
the recurrence relations, which appear to be related to the 
universal functions describing soft gluon emission, at least for the 
case discussed here.

In general, we envisage that the logarithmic parts of one-loop QCD
amplitudes will be obtainable from unitarity or related methods,
while the rational function parts will be calculable using recursive
methods.  We can think of all-gluon QCD amplitudes as being composed
of three distinct terms: an $\NeqFour$ supersymmetric amplitude; a
multiple of the contribution of an $\NeqOne$ supersymmetric chiral
multiplet; and a multiple of the contribution of a colored scalar
circulating in the loop.  (There is a similar type of decomposition
for amplitudes with external fermions~\cite{QQGGG}.)
The first two terms can be computed from
their four-dimensional cuts, for which in turn the computation of
massless four-dimensional helicity amplitudes suffices.  
Many such amplitudes have already been computed, in both
the $\NeqFour$~\cite{Neq4Oneloop,BCF7,NeqFourSevenPoint,NeqFourNMHV}
and $\NeqOne$ cases~\cite{Neq1Oneloop,OtherNeqOne,DunbarNeq1}

The scalar term, which is non-supersymmetric, can also be computed via
unitarity, but the computation would require tree amplitudes with the 
scalar propagating in the full $D=4-2\e$ 
spacetime~\cite{UnitarityMachinery}, which are more cumbersome
for large numbers of external legs. 
We can, however, think of splitting the scalar contribution up 
further into those terms which can be computed from
the four-dimensional cuts, and a remainder.  Indeed, the cut-containing
parts of the scalar loop amplitudes with two negative-helicity gluons 
and $(n-2)$ of positive helicity have already been 
computed~\cite{Neq1Oneloop,BBSTNeq0}. To the terms with
discontinuities such as dilogarithms and logarithms,
it is convenient to add certain rational terms which remove spurious
(unphysical) singularities~\cite{GGGGG}.  
For example, the function $\ln[(-s_1)/(-s_2)]/(s_1-s_2)^2$
generically appears in one-loop amplitudes, and has a singularity 
as the momentum invariant $s_1$ approaches $s_2$, which is not associated with
any physical factorization.  This singularity can be removed by replacing
\be 
{ \ln\Bigl( {-s_1 \over -s_2} \Bigr) \over (s_1-s_2)^2 }
\to { \Ll_1\Bigl( {-s_1 \over -s_2} \Bigr) \over s_2^2 } \,,
\ee
where $\Ll_1(r) \equiv (\Ll_0(r)+1)/(1-r)$, 
$\Ll_0(r) \equiv \ln(r)/(1-r)$.
Subtracting from the full amplitude all the cut-containing terms, 
and those rational functions linked to them by spurious singularities,
defines a `pure-rational' remainder.
This function may be computed by on-shell recurrent techniques as 
done at tree level by BCFW~\cite{BCFW}, and at one loop in the present paper.
To analyze the factorization properties, that is the residues at
the poles in $z$, it may prove useful to think of the pure-rational 
remainder as the difference between the scalar loop amplitude
in $D=4-2\e$ and that in $D=4$.  Terms in the loop integral
which are proportional to the $(-2\e)$ components of the loop 
momentum are well-behaved in the infrared.  For this reason, 
complications (discontinuities) in amplitude factorization which
are associated with infrared divergences~\cite{BernChalmers} 
should be absent from the appropriate differences of amplitudes.

One may wonder whether one can push this approach further.  Indeed, it
seems quite likely that it will apply at tree level to a wide variety
of field theories, including scalar field theories or gauge theories
with massive fermions or scalars, and indeed even to
spontaneously-broken gauge theories. Another question is to what
extent it can be applied beyond tree level, and indeed beyond the
rational terms at loop level as discussed above.  At loop level, the
amplitude will now have branch cuts in $z$; but in general, we may
still expect it to fall off at infinity, allowing $A(0)$ to be
analyzed in terms of simpler (lower-point) amplitudes.
We are optimistic that these techniques will open the door
to many new computations of Standard Model scattering amplitudes.


\section*{Acknowledgments}

We thank Ed Witten, Peter Svr\v{c}ek and Valya Khoze
for helpful conversations and comments,
and Freddy Cachazo for a relevant comment.  We also thank
the Mathematical Institute at Oxford University, where this work was
begun, and Academic Technology Services at UCLA for
computer support. 


\end{document}